\documentclass[aps,prl,twocolumn,groupedaddress,a4paper]{revtex4}
\pdfoutput=1
\usepackage{graphicx}
\usepackage{amssymb}
\usepackage{amsmath}
\usepackage{amsfonts}
\usepackage{bm}
\usepackage{color}
\usepackage{times}
\newcommand{\ket}[1]{\ensuremath{\left|{#1}\right\rangle}}

\newcommand{\oper}[1]{\boldsymbol{#1}}

\begin{document}


\title{Experimental Observation of Quantum Chaos in a Beam of Light}

\author{Gabriela B. Lemos}
\email[]{gabriela.barreto.lemos@univie.ac.at}
\affiliation{Instituto de F\'{\i}sica, Universidade Federal do Rio de
Janeiro, Caixa Postal 68528, Rio de Janeiro, RJ 21941-972, Brazil}
\author{R. M. Gomes}
\affiliation{Instituto de F\'{\i}sica, Universidade Federal do Rio de
Janeiro, Caixa Postal 68528, Rio de Janeiro, RJ 21941-972, Brazil}
\author{S. P. Walborn}
\affiliation{Instituto de F\'{\i}sica, Universidade Federal do Rio de
Janeiro, Caixa Postal 68528, Rio de Janeiro, RJ 21941-972, Brazil}
\author{P. H. Souto Ribeiro}
\affiliation{Instituto de F\'{\i}sica, Universidade Federal do Rio de
Janeiro, Caixa Postal 68528, Rio de Janeiro, RJ 21941-972, Brazil}
\author{F. Toscano}
\affiliation{Instituto de F\'{\i}sica, Universidade Federal do Rio de
Janeiro, Caixa Postal 68528, Rio de Janeiro, RJ 21941-972, Brazil}

\begin{abstract}
The manner in which unpredictable chaotic dynamics manifests itself in quantum mechanics is a key question in the field of quantum chaos. Indeed, very distinct quantum features can appear due to underlying classical non-linear dynamics. Here we observe signatures of quantum non-linear dynamics through the direct measurement of the time-evolved Wigner function of the quantum Kicked Harmonic Oscillator, implemented in the spatial degrees of freedom of light. Our setup is decoherence-free and we can continuously tune the semiclassical and chaos parameters, so as to explore the transition from regular to essentially chaotic dynamics.  Due to its robustness and versatility, our scheme can be used to experimentally investigate a variety of non-linear quantum phenomena. As an example, 
we couple this system to a quantum bit and experimentally investigate the decoherence produced by regular or chaotic dynamics.
\end{abstract}



\maketitle
\section*{Introduction}
Chaotic classical systems have the characteristic trait of being extremely  sensitivity to initial conditions.  This behavior, together with the experimental imprecision of the initial conditions, cause these deterministic systems to be inherently unpredictable.  The field of quantum chaos addresses the question as to how classical chaotic dynamics 
manifests itself in quantum mechanics.  In addition to fundamental questions concerning the correspondence principle and the 
 classical limit of quantum mechanics, a number of intriguing quantum-dynamical features have been unravelled. 
Prominent examples are dynamical localization \cite{Izrailev},  the 
quantum suppression of classical diffusion, and the enhancement of the tunneling rate in the presence of chaos in the corresponding classical dynamics 
\cite{steck2001,Hensinger2001}.
These phenomena have been observed in several physical systems 
\cite{raizen_pre1999,kr-exp0, steck2001,Hensinger2001, nature-chaos,Chabe2008,Sadgrove_review,Ryu2006,Fischer2000,Schwartz2007}.
\par 
The simplest and most widely studied systems that present manifestations of classical chaos in their quantum dynamics are periodic time-dependent Hamiltonian (Floquet) systems \cite{casati-chirikov}.
The quantum evolution up to discrete time $t=nT$ is described by the quantum map
\begin{equation}\label{eq:evolucao}
 |\psi(n)\rangle=\oper{U}^n |\psi(0)\rangle,
\end{equation}
where $n$ is an integer and the Floquet operator $\oper{U}$ describes the unitary quantum evolution in the time period $T$.  In addition to extensive study from a theoretical viewpoint, the phenomena arising in these maps have been observed in experiments with atoms \cite{kr-exp0, steck2001,Hensinger2001, nature-chaos,Chabe2008,Sadgrove_review,Schlunk2003}, Bose-Einstein condensates \cite{Ryu2006}, and photonics lattices \cite{Schwartz2007}. There have been a few theoretical proposals to realize quantum chaotic maps using paraxial optics \cite{Prange,berry,baker-osorio} and, in fact, dynamical localization has been observed in an optical field sent through a sequence of phase gratings \cite{Fischer2000}.    
However, the realization of the quantum kicked harmonic oscillator (KHO), a paradigm of quantum  non-linear dynamics with non-KAM (KolmogorovÐArnoldÐMoser) behavior, and a model for charges moving in time-dependent fields~\cite{zaslavsky}, electronic transport in semiconductor lattices~\cite{fromhold,fromhold1}, and trapped ions in a periodic laser field \cite{cirac-zoller}, is still outstanding. 
\par
The quantum KHO is described by the map \eqref{eq:evolucao}, where the iteration operator is 
\begin{equation} \label{eq:evolucao-kho}
\oper{U}_{\rm{KHO}} = \oper{R}_{{\alpha}}\oper{V}_K.
\end{equation} 
The operator $\oper{R}_{{\alpha}}$ describes the evolution of a quantum harmonic oscillator parameterized by $\alpha=\omega T$, where 
$\omega$ is its frequency, and $T$
the interval between the periodic perturbations. $\oper{V}_K$ describes a periodic perturbation corresponding to a potential $K\cos(Q + \phi)$, where $\phi$ is a phase and $Q$ is the dimensionless position variable defined below.
\par
The dimensionless position and momentum coordinates of a particle of mass $m$ submitted to the KHO evolution are defined as $Q=\nu q$ and $P=\nu p/(m\omega)$ respectively, where $q$ and $p$ are the position and momentum of the particle and $\nu$ is the spatial frequency of the kick.  The dimensionless operators obey $[\oper Q,\oper  P]=i\hbar_{\rm eff}$, where the effective Planck constant is $\hbar_{\rm eff}\equiv\nu^2\hbar/m\omega$.
\par
This  model can present a rich variety 
of intriguing quantum-dynamical phenomena\cite{artuso_prl1992,carvalho_buchleitner2004}.   In the so-called quasicrystal condition 
and also for irrational values of $\alpha$, a quantum localization, similar to that extensively studied and observed in the kicked rotor model, can appear \cite{Borgonovi1995}.  For the crystal condition where $\alpha \in \{ \pi/3, \pi/2, 2 \pi/3, \pi, 2 \pi \}$, the quantum system can present diffusion in energy 
for any value of $K$. In this case, the stroboscopic phase space of the corresponding classical system is characterized by the appearance of a 
``stochastic web" associated to the chaotic behavior, with 
periodic regions corresponding to essentially regular dynamics in between. 
Stochastic webs are typical structures of systems
with non-KAM  behavior, where chaotic dynamics appears even for an arbitrary small perturbation (in our case the kicks) \cite{zaslavsky}, with 
 have many applications \cite{fromhold1}. The 
size of the web and the perturbation of the regular dynamics 
inside the periodic regions are governed by the intensity of the perturbation $K$.
For $K< 1$ the size of the web is considerably small, as is the perturbation
of the regular dynamics inside the periodic regions.   When $K = 2$ the KHO can be considered a weakly chaotic system. 
\par 
Here we implement the quantum KHO dynamics in the spatial degrees of freedom of the photons of a monochromatic paraxial light. We observe the non-linear dynamics through direct measurement of the optical Wigner function.
Controllable parameters adjust the system from regular to chaotic dynamics, as well as the effective Planck constant, associated to the quantum-classical transition.  Our scheme is decoherence free and can be employed in a variety of studies of non-linear quantum systems, which we illustrate by investigating the decoherence induced by our system on a qubit.
\par
\begin{figure}
\begin{center}
    \includegraphics[width=8cm]{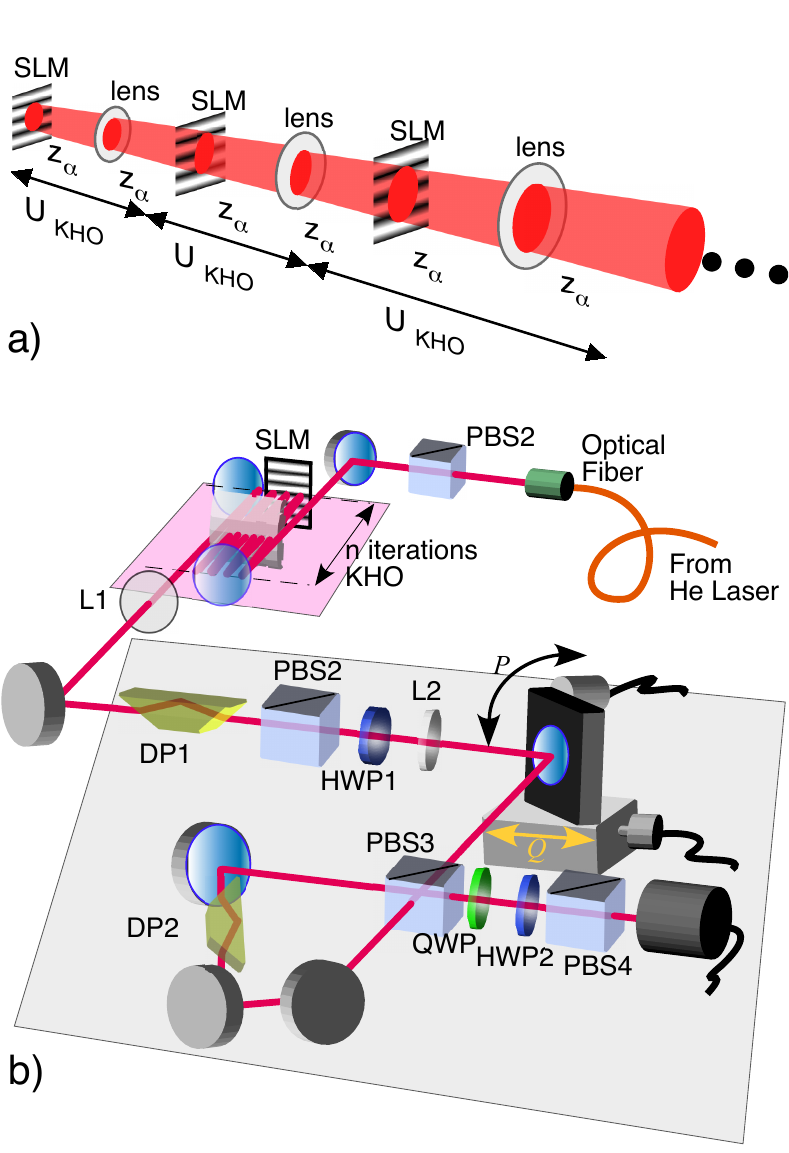}
    \end{center}
  \caption{   
\textbf {Optical scheme and experimental setup.} \textbf{a} Illustration of
an optical system to implement the quantum KHO. \textbf{b} Experimental
setup for the implementation of the quantum KHO
and measurement of the spatial Wigner function.  The output of the HeNe laser is horizontally polarized using a polarizing beam splitter (PBS1) and reflected into $n$ iterations of the KHO operation.  Each iteration consists in a ``kick", corresponding to a phase imprinted on the field with the spatial light modulator (SLM), and harmonic evolution, implemented with sections of free propagation and a cylindrical lens.  The beam is reflected back and forth $n$ times onto the SLM. Lenses L1 and L2 map the final state at transverse plane $z_0$ onto the entrance
of a Sagnac interferometer. Before entering the interferometer, a Dove 
prism (DP1) is used to perform a $90^\circ$ spatial rotation of the beam profile,
and a polarizing beam splitter (PBS2) and the half-wave plate (HWP1) are
used to balance the intensities of the vertical and horizontal polarization components.
The interferometer is used for direct measurement of the optical Wigner 
function. Each phase space point of the Wigner function is obtained from the interference 
between horizontal ($H$) and vertical ($V$) polarization components of
the beam emerging from the interferometer. A Dove prism (DP2) inside 
the interferometer is used to spatially rotate the counter-propagating 
$H$ and $V$ components of the field. Translation and tilting of the input 
mirror are used to select the phase space point $(Q,P)$ to be 
measured. A quarter-wave plate (QWP), a half-wave plate (HWP2) and a 
polarizing beam splitter (PBS4) are used to perform the required polarization measurement, and a large aperture power meter
is used to measure the beam intensity. 
See the Methods section for more information.}
\label{fig:setup}
  \end{figure}
\section*{Results}

\begin{figure*}
  \begin{center} 
\includegraphics[width=16cm]{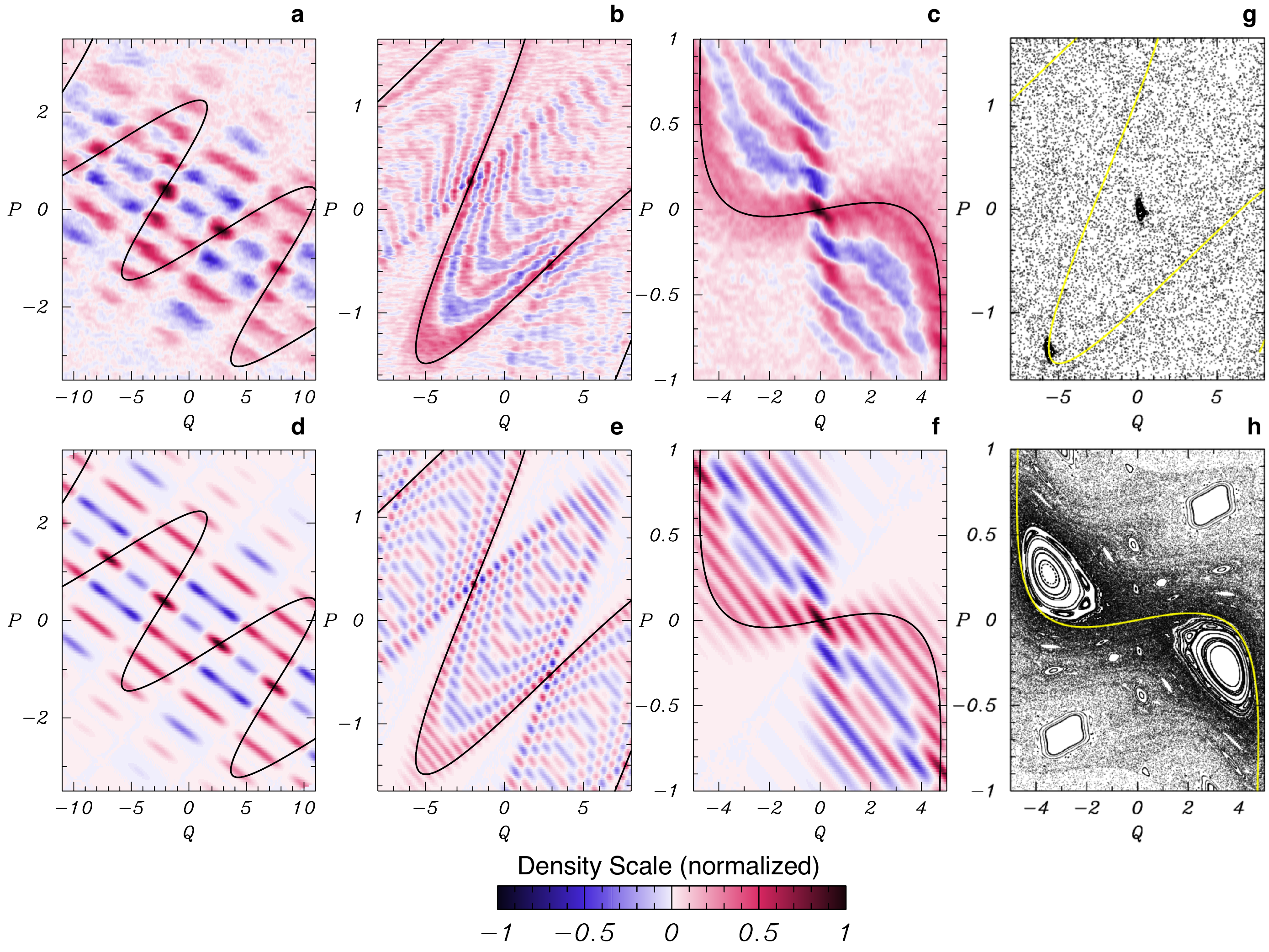}
\end{center}
  \caption{{\bf Quantum and classical phase space for a single iteration of the KHO map.}  In the top row from  {\bf (a)} to {\bf (c)}, density plots of the directly measured optical Wigner distributions for $n=1$ iterations of the KHO map; in the bottom row
from {\bf (d)} to {\bf (f)},  the corresponding numerical predictions. For {\bf (a)} and {\bf (e)}, the relevant parameters are $K=7.4$, $\hbar_{\rm eff}=4.72$ and 
$\phi=1.45\pi$; 
for {\bf (b)} and 
{\bf (f)}, $K=7.4$, $\hbar_{\rm eff}=0.9$ and  $\phi=1.4\pi$;
and for {\bf (c) } and {\bf (g)}, 
$K=2$, $\hbar_{\rm eff}=0.9$ and $\phi=0$.
In all cases the harmonic evolution parameter is $\alpha=\pi/3$.
The black line outlines the phase space manifold that is the skeleton of both the quantum and the classical distributions corresponding to the KHO Hamiltonian
(see details in the text).  
The associated stroboscopic phase space evolution of the classical map
 is illustrated 
in  {\bf (g)} and {\bf  (h)} corresponding to the cases in {\bf (b) } and {\bf (c)} respectively (the skeleton manifold of the evolved state is the yellow line).
The theoretical plots (including the stroboscopic phase space) are obtained from 
the evolution of the KHO corrected by a linear transformation (equal for all the plots)
that take into account the spurious evolution due to technical imperfections and small alignment errors of the optical elements (see the Methods section).} 
     \label{wig1}
  \end{figure*}
\begin{figure*}
  \centering 
 \includegraphics[width=16cm]{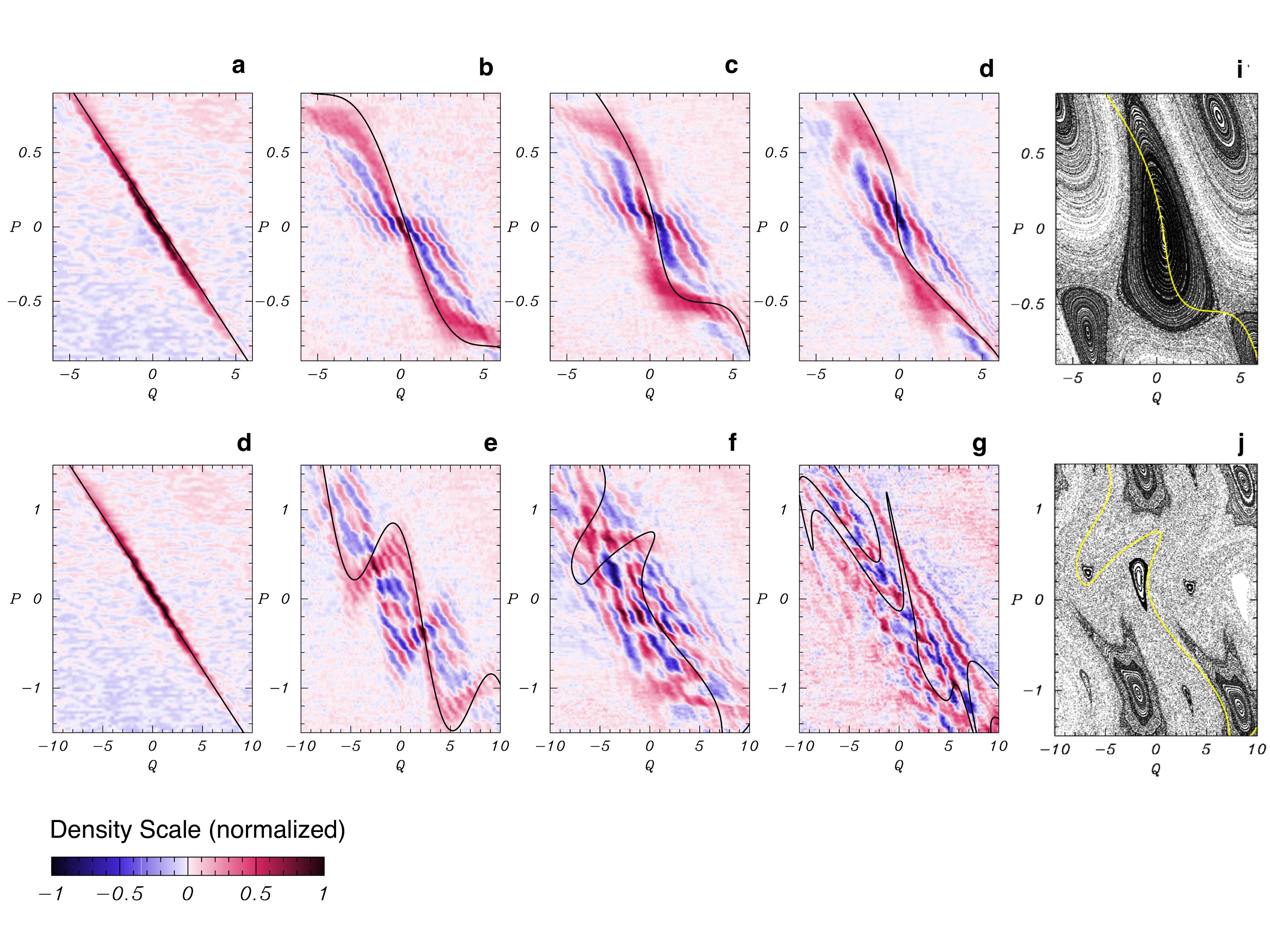} 
  \caption{ {\bf Experimentally obtained Wigner distributions corresponding to the first three iterations of the KHO map. } 
  The initial Gaussian state is shown in plot {\bf (a)} ( $\hbar_{\rm eff}=0.42$) and in {\bf (e)} ($\hbar_{\rm eff}=0.9$). 
  In the top row from  {\bf (b)} to {\bf (d)}, the case of almost regular dynamics, with $K=0.75$, $\hbar_{\rm eff}=0.42$ and $\phi=0.44\pi$.  
In the second row from {\bf (f)} to {\bf (h)} the case of weak chaos regime, with $K=2$,  $\hbar_{\rm eff}=0.9$ and $\phi=1.33\pi$. 
In all cases $\alpha = 2\pi/3$.  
The corresponding stroboscopic classical phase space are shown in {\bf (i)} ($K=0.75$) and{\bf (j)} $K=2$.
The black lines correspond to the skeleton of both the quantum and the classical distributions corresponding to the KHO (obtained in the same way as in Fig. \ref{wig1}). 
In the bottom row, the recognizable positive skeleton that the Wigner function shares with the classical probability distribution (obtained evolving classically the initial distribution of the
Gaussian state) quickly disappear due to the chaotic dynamics ($K=2$) in comparison with the essentially regular dynamics ($K=0.75$) pictured in the top row.
}
    \label{wig3}
  \end{figure*}

\subsection*{Optical implementation of the chaotic quantum map}
We implement an optical version of the operator \eqref{eq:evolucao-kho} in the spatial degrees of freedom of monochromatic paraxial light based on the isomorphism between the paraxial wave equation and the Schr\"odinger equation (see \cite{stoler,shamir82,nienhuis93}
and Supplementary Note 1).   
 The light beam is sent $n$ times through a combination of optical elements designed to implement the operator $\oper{U}_{\rm{KHO}}$,  as illustrated in Fig.\ref{fig:setup}(a).  The transverse position in the near and far-field correspond to the position and momentum of the photons in the beam, and are analogous to the transverse position $q$ and transverse momentum $p$ of a quantum particle. The instantaneous ``kick" perturbation, 
\begin{equation}
\oper{V}_{K} = e^{-\frac{ i}{ \hbar_{\rm eff}} K \cos \oper{Q}},
\label{eq:kick}
\end{equation}
is produced using a Holoeye spatial light modulator (SLM). 
The SLM imprints a programmable phase  $\exp[i f(x,y)]$ on an optical beam, and thus can be used in the implementation of many dynamical maps.
We define $Q=\nu q$ as the dimensionless version of the near-field variable, $q$ (see Supplementary Note 2). The parameter $\nu$ is the spatial frequency of the cosine function, $K$ is the kick strength, and the effective Planck constant $\hbar_{\rm eff}$ will be defined below. 
\par
The harmonic evolution operator $ \oper{R}_{\alpha}$   
produces a phase space rotation that  
 is equivalent to a Fractional Fourier Transform (FRFT) of order $\alpha$~\cite{ozaktas}, 
\begin{equation}
\oper{R}_{\alpha} = e^{i \alpha/2} e^{\frac{i}{ \hbar_{\rm eff}} \alpha(\oper{Q}^2+\oper{P}^2)/2},
\label{eq:ho}
\end{equation}  
which can be implemented using a lens of focal length $f$ placed between two sections of free-space of distance $z_\alpha=2f\sin^2(\alpha/2)$.
The dimensionless momentum variable is $P=\nu f^\prime \theta$, where $\theta$ is the angle of the paraxial ray and $f^\prime \equiv f \sin \alpha$. The dimensionless operators $\oper{Q}$ an $\oper{P}$ obey the commutation relation~\cite{stoler,shamir82},  
\begin{equation}\label{eq:comutacao}
 [\oper Q,\oper  P]=i\hbar_{\rm eff},
\end{equation}
where  $\hbar_{\rm eff} = \nu^2 f^\prime/k$ is the
dimensionless effective Planck constant  (see Supplementary Note 2).  
\par
It is straightforward to manipulate all the relevant parameters in the dynamics of the KHO:
By changing the order $\alpha$ of the FRFT (harmonic evolution between kicks), the amplitude $K$ of the cosine phase implemented with the SLM, and the spatial frequency $\nu$ of this phase (effective Planck constant).
\par
The experimental setup is illustrated in Fig.\ref{fig:setup}(b).
To characterize the chaotic dynamics, we do a point-by-point direct measurement of the optical Wigner function~\cite{bastiaans,ozaktas} using an interferometric method~\cite{walmsley-wigner}. Details of the full experimental setup are given in the methods section, and in the Discussion section we show how this scheme can be used as a building block to implement long-time dynamics.
\par
\subsection*{Observing quantum signatures of chaos in phase space}
Figs. \ref{wig1}(a-c) show three experimental Wigner distributions for a single iteration ($n=1$) of the quantum map \eqref{eq:evolucao-kho} applied to a squeezed Gaussian state $\vert \Psi(0)\rangle$ centered at the phase space origin (shown in Figs.\ref{wig3}(a) and (e)).
The theoretical Wigner distributions are shown for comparison in Figs. \ref{wig1}(d-f).  All three cases correspond to the harmonic evolution $\alpha=\pi/3$.   
The classical dynamics of the KHO map (controlled by the kick amplitude $K$) are illustrated  with the usual stroboscopic kick-to-kick map in Figs. \ref{wig1}
(g) and (h). 
Figs. \ref{wig1}(a, d) and (b, e) show results for the kick amplitude $K=7.4$
that corresponds classically to essentially chaotic dynamics 
(see Fig. \ref{wig1} (g)). Figs. \ref{wig1}(c, f) show results for $K=2$ that
correspond to a mixed classical dynamics (see Fig. \ref{wig1} (h)).
\par
Extended quantum states (with uncertainty $\Delta \oper{Q}\Delta \oper{P} >>
\hbar_{\rm eff}/2$) typically exhibit a fine oscillatory structure in their Wigner
function, known as sub-Planck structure \cite{zurek-nature}, which saturate at a  scale  
$\sim\hbar_{\rm eff}^2/\Delta \oper{Q}\Delta \oper{P}$.  Thus, for an almost fixed extension in phase space,  
the wavelength of the oscillatory pattern should decrease with $\hbar_{\rm eff}$. 
This can be observed comparing Figs. \ref{wig1} (a,d), where $\hbar_{\rm eff}=4.72$, with Figs. \ref{wig1} (b,e)  where $\hbar_{\rm eff}=0.9$ (in both cases $K=7.4$).
\par
The Wigner function of some extended states, like energy eigenstates in integrable systems, have a classical manifold as their support. Every pair of localized regions of the Wigner function on this support can interfere creating an oscillatory pattern in the middle of a chord that joins the localized regions, 
similar to the Wigner function of a superposition of two coherent states 
(a Schrodinger cat-like state).
When the classical dynamics is sufficiently non-linear, the evolution of even highly localized Gaussian states are supported by a phase space manifold that evolves classically \cite{Maia2008,Schubert2012}. 
This is also the skeleton of the classical probability distribution associated with the initial Gaussian state whose evolution is determined by the Liouville equation of the classical system.
\par
In all the plots of Fig. \ref{wig1} a curve (yellow in 
(g) and (h), black in the rest) indicates the classical manifold that is the support of the Wigner function. For our initial squeezed Gaussian state this is a straight line, as in Figs. \ref{wig3}
(a) and (e).
The interference pattern appears once 
the stretching and folding of the classical manifold begins due to 
the non-linear classical dynamics. 
Eventually, some portion of the  interference pattern falls 
over the classical support,  and the positive 
skeleton begins to disappear. This occurs very quickly when the underlying dynamics is chaotic. This can be seen in Fig. \ref{wig3}, which shows the Wigner function of the first three iterations of the KHO map \eqref{eq:evolucao-kho} for the same initial state $\vert \Psi(0)\rangle$ (shown in (a) and (e)),
when $K=0.75$ in (b) to (d)  and when $K=2$ in  (f) to (g).
The non-linear regular classical dynamics of a stability island around the origin for $K=0.75$ is shown in the stroboscopic phase space plot (i), and the weak chaotic dynamics 
for $K=2$ in plot (j).
\par

\subsection*{The quantum KHO as a decohering environment}
\begin{figure}
  \centering 
  \includegraphics[width=8cm]{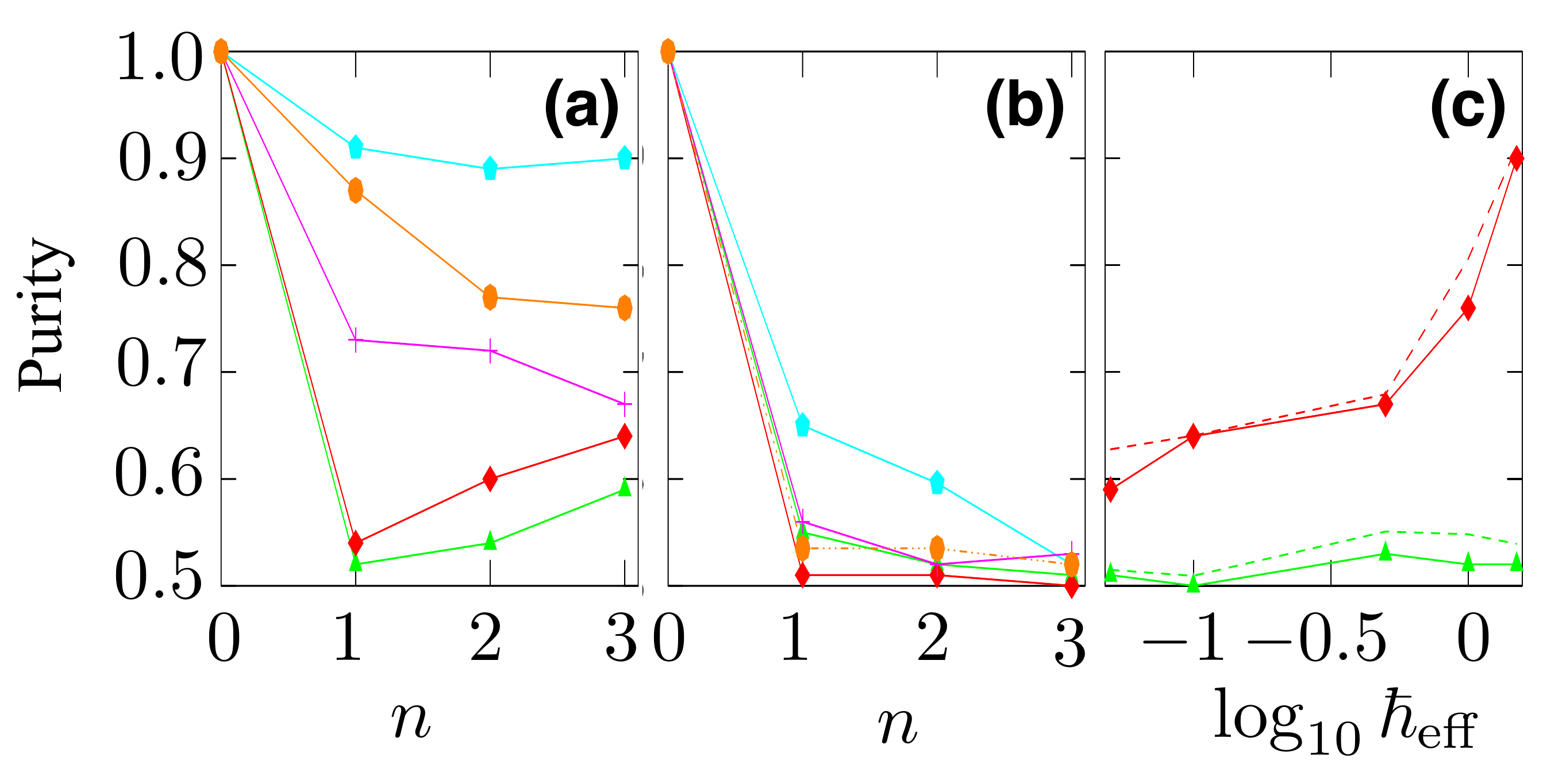} 
   \caption{{\bf Purity loss of the diagonal polarization state for regular and chaotic dynamics of the environment.} 
   Purity as a function of the number of  KHO iterations, $n$, for {\bf (a)} regular dynamics ($K=0.5$),  and {\bf (b)} chaotic dynamics ($K=2$). 
   In both figures $\alpha=2\pi/3$, and $\hbar_{\rm eff} =0.05$ (green triangles), $0.1$ (red diamonds), $0.5$ (purple crosses), $1.0$ (orange circles), and $1.5$ (blue polygons). {\bf (c)} the final purity  as a function of  $\hbar_{\rm eff}$ for $n=3$ KHO iterations, with $K=2$ (green triangles) and $K=0.5$ (red diamonds). The dashed lines show the predictions given by numerical simulation of the composite qubit-KHO system. The solid lines joinling the experimental points are only to assist visualization.}
     \label{pure}
  \end{figure}
As an example of the utility and versatility of
our approach, we measured the loss of coherence
in a polarization qubit coupled to the quantum KHO, which acts as a decohering environment. By 
taking advantage of the fact that the SLM only imprints a phase on the horizontal polarization component, a polarization dependent evolution, KHO or simple harmonic oscillator (SHO), is 
implemented in the spatial degrees of freedom  (see Methods section). This corresponds to a dephasing-type interaction between qubit and environment. In this case, the off-diagonal elements of the qubit density matrix are suppressed by a factor $f=\vert\langle\Psi(0)|\left(\oper U_{\rm SHO}^n \right)^\dagger\oper U_{\rm KHO}^n|\Psi (0)\rangle\vert$, where $|\Psi (0)\rangle$ is the initial state of the environment.
Thus, for an initial state of the qubit in the 
equatorial plane of the Bloch sphere, the temporal behavior of its purity
is given by $ (1+|f|^2)/2$.
The quantity $f$ can be seen as a  fidelity amplitude, which has  been extensively studied in the field of quantum chaos \cite{haug2005, gorin2006, Lemos2011}.
\par
In general,  theoretical studies predict an initial decay of $f$ before saturation  \cite{haug2005,gorin2006,Lemos2011}.
For fully chaotic underlying classical dynamics, $f$ presents an exponential decay with different decay rates depending on the perturbation regime \cite{haug2005, gorin2006}.
On the other hand, for regular dynamics, the decay of $f$ is not generic and depends strongly on the localization of the  initial state in the classical phase space \cite{haug2005, gorin2006}. 
For initial states well localized in a stability island long time oscillations
with revivals are expected, where the oscillations can be understood in terms of the classical frequencies included in $|\Psi (0)\rangle$.
The temporal mean value of $f$ decreases with $\hbar_{\rm eff}$
so, due to revivals, the size of the fluctuations increase in the semiclassical regime. 
When the underlying dynamics of the environment is chaotic the temporal mean value of $f$ and its fluctuations are inversely
proportional to the effective Hilbert space dimension of the
environment \cite{Lemos2011} and therefore tend to zero in the semiclassical limit ($\hbar_{\rm eff}\rightarrow 0$). In this limit the qubit becomes maximally entangled to the environment, so its purity$\longrightarrow 1/2$.
\par
In our experiment,  an incoming beam was prepared in a linear diagonal polarization state. Figs. \ref{pure}(a) and (b), shows the purity of the polarization state as a function of the number of iterations of the KHO map.  Fig. \ref{pure} (a) is for essentially regular dynamics ($K=0.5$), and Fig. \ref{pure}(b) shows the case in which the KHO has chaotic dynamics ($K=2$).  
The initial state $\vert \Psi(0)\rangle$ is analogous to the one shown in 
Fig. \ref{wig3} (a) and (g) and in the case when $K=0.5$ is localized in an stability
island around the origin (not shown).
 Fig. \ref{pure}(c) shows  the purity  for $n=3$ kicks as a function of $\hbar_{\rm eff}$ for $K=2$ (green triangles) and $K=0.5$ (red diamonds). The dashed lines are the predictions given by numerical simulation of the composite system.
Although the number of kicks are few,  one observes  a general behavior compatible with the theory described above.
In the case of chaotic dynamics ($K=2$), rapid loss of purity occurs for all values of $\hbar_{\rm eff}$ attaining a saturation with very small fluctuations, indicating  that the polarization  state becomes nearly maximally entangled (purity $=1/2$)
for decreasing values of  $\hbar_{\rm eff}$ (see Fig. \ref{pure}(c)). Hence, the equilibrium state of the qubit is a totally mixed state.
The total loss of coherence in the polarization state here is due to the underlying classical chaotic dynamics in the spatial degrees of freedom of the beam.  
On the other hand, for regular dynamics ($K=0.5$), because $|\Psi (0)\rangle$
is almost completely localized in an stability island, 
we observe what appears to be the beginning of oscillations for different values of $\hbar_{\rm eff}$, with a revival of the polarization state purity (for all values of $\hbar_{\rm eff}$). The value of the purity at its minimum value goes to $1/2$ in the semiclassical limit (see Fig. \ref{pure}(c)) indicating that the temporal mean value of 
$f$ goes to zero when $\hbar_{\rm eff}\rightarrow 0$. This is compatible with the typical  large fluctuations of  the fidelity amplitude $f$ in the semiclassical regime for the case of regular dynamics when the initial state is localized in a stability island.
 \par
  
\section*{Discussion}

\begin{figure}
\includegraphics[width=8cm]{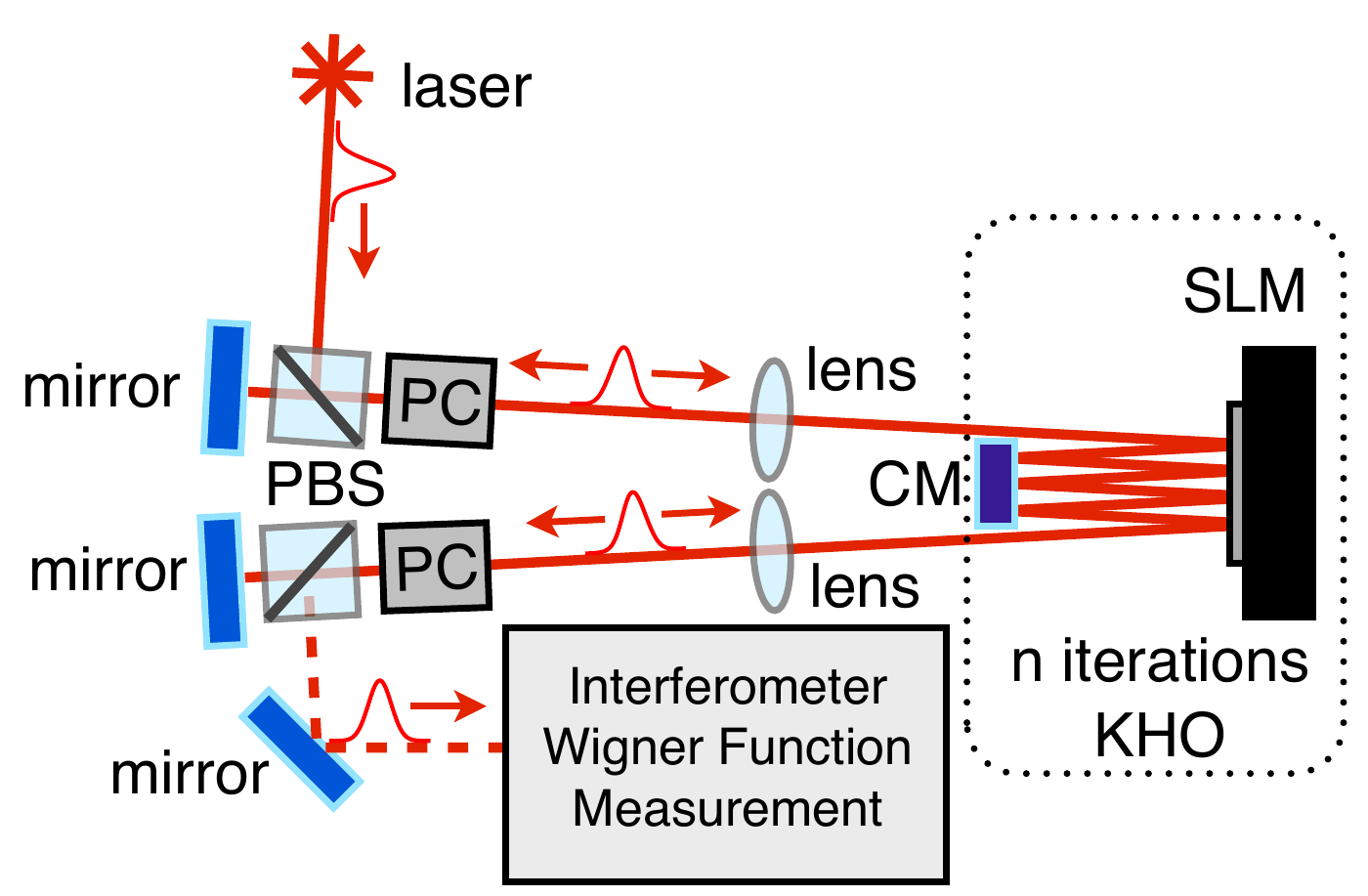}
\caption{{\bf Scaling up to a large number of kicks.} The $n$ iterations of the KHO operator can be used as a building block for $N>>1$ kicks. An optical pulse is switched into the setup using a Pockels cell (PC).  The $n$ KHO iterations are performed using a cylinder mirror (CM) and spatial light modulator (SLM). After a number of KHO evolutions, the pulse can be switched out of the setup and into the interferometer for direct measurement of the Wigner function.The ultimate limit to the number of iterations that can be performed depends upon the losses in the optical system (see the Supplementation Discussion).}
\label{fig:khoprop}
\end{figure}

The optical KHO setup reported here can be used as a building block to implement a large number of iterations of the KHO operator.  Fig. \ref{fig:khoprop} illustrates an experimental scheme that can be used to implement $N>>1$ kicks of the KHO.  A pulse from a vertically-polarized laser is reflected from a polarizing beam splitter (PBS), and the polarization is rotated to the horizontal direction by a Pockels cell (PC).  The laser is sent to the SLM, and $n$ iterations of the KHO operator are implemented, in the same manner as reported in the Results section.  To minimize losses from multiple optical components, a single cylindrical mirror (CM) can be used in the place of the cylindrical lens and plane mirror in Fig. \ref{fig:setup}.  After $n$ iterations, the output light is sent through a second PC, which can be used to switch the pulse out of the setup for measurement.  The measurement system is the interferometer used for direct measurement of the spatial Wigner function.  If the PC is left inactive, the pulse is reflected from the mirror and backwards through the KHO operation, resulting in another $n$ iterations.  The lenses are chosen with focal length equal to half the distance between the SLM and mirrors, so that two consecutive optical Fourier transforms are performed.  The overall result is an imaging system, up to a reflection.  In this way, the output state from each set of $n$ iterations is mapped onto the input state for the next set of $n$ iterations.  Since the kick operator \eqref{eq:kick} and harmonic evolution \eqref{eq:ho} are symmetric around the origin, the reflections can be absorbed into the definition of the coordinate system.  The pulse is sent back to the first PC, which remains inactive, resulting in reflection of the pulse at the mirror, and transmission back into the $n$ KHO iterations. In this way, it is possible to implement a sequence of $n\times 2n \times 2n\cdots$ kicks and switch the output state into the interferometer for measurement after the first $n$ iterations, or at intervals of $2n$ after that.  We note that the SLM allows us to control the number of kicks $n$ per iteration by programming whether the kick-phase or a quadratic phase is imprinted on the field.  The quadratic phase can be used to implement or undo the harmonic evolution.  
\par
The ultimate limit to the number of iterations that can be performed depends primarily upon the losses in the optical system.  These arise predominantly from the SLM, lenses and mirrors used in the $n$ iterations of the KHO.  The total output intensity after $n$ kicks of the KHO can be written
\begin{equation}
I_{\mathrm{out}} = I_{\mathrm{in}}t_{o}(t_{l} t_{\mathrm{SLM}})^n,
\label{eq:loss}
\end{equation}
where $t_o$ is the combined transmission coefficient of the optical elements outside the KHO evolution, $t_l$ is the combined transmission coefficient of the lenses and mirrors that implement the harmonic evolution between kicks, $t_{\mathrm{SLM}}$ the transmission coefficient of the SLM, and $I_{\mathrm{in}}$ the input intensity of the laser beam.   In the Supplementary Discussion, we estimate that with current technology it should be possible to perform about $n \sim 100$ kicks.   
\par
In conclusion, our experiment allows for the study of the dynamics of a non-relativistic quantum system using an intense classical laser beam due to the analogy between quantum mechanics and classical wave mechanics.   A possible next step is to study the chaotic evolution of entangled photons.  The optical realization of non-linear quantum dynamics should prove invaluable in the experimental investigation of quantum chaos, decoherence, and the quantum-classical boundary. 


\section*{Methods}\label{Methods}
\subsection*{Experimental setup}
The complete experimental setup is illustrated in Fig.\ref{fig:setup}(b). A 632.8nm He-Ne laser is coupled to a single-mode optical fiber. This defines a Gaussian light beam as the initial state $\ket{\psi (0)}$ which is then evolved by the quantum KHO propagator \eqref{eq:evolucao-kho} in one dimension of the transverse spatial degrees of freedom.   The state is sent through $n$ iterations of the KHO operator $U_{KHO}$ by reflecting $n$ times between the SLM and a mirror.  
A polarizing beam splitter (PBS1) is used to polarize the beam parallel to the active axis of the SLM display.  The harmonic evolution between kicks corresponds to propagation in the $\alpha$-order FFT system, which consists of free space propagation and the cylindrical lens.
For practical reasons, we actually implement two consecutive FFTs of order $\alpha/2$, before each incidence on the SLM. The focal length
of the cylindrical lens is $f = 150$mm, and the free space propagation
length is $z = 75$mm, so that $\alpha = \pi/3$ between two consecutive kicks (see Fig.\ref{fig:setup}(b)). Since the entire KHO evolution is made with optical elements that act only in one spatial dimension, the perpendicular direction evolves according to free space propagation
(see the Supplementary Discussion
for details and a discussion of the accessibility of long-time dynamics). 
\par In order to measure the Wigner function, two spherical lenses ($L1$ and $L2$) of focal length $f=350$ mm are used to map the output state of the KHO system $\vert\psi (n)\rangle$ from the transverse plane at position $z_{0}$  to the input mirror of the Sagnac interferometer. 
A Dove prism (DP1) tilted at a $45^\circ$ angle is used to swap the horizontal and vertical coordinates, because for convenience the quantum kicked Hamiltonian is implemented in the vertical axis, while the Wigner measurement is {performed in the horizontal}.  The second PBS (PBS2)
and the first half wave plate (HWP1) are used to keep the beam linearly polarized before entering the Sagnac interferometer. 

\subsection*{Measurement of the optical Wigner function}
  The method used to {directly} measure the optical Wigner function is an interferometric scheme proposed in reference \cite{walmsley-wigner}.  The interferometer is illustrated in Fig. \ref{fig:setup}.    
  The displacement and tilting of a steering mirror ($M1$) at the entrance of a three-mirror Sagnac interferometer displaces the optical field by $Q$ and changes its 
  direction of propagation by $P$ (which in the paraxial approximation corresponds to the addition of a phase). This produces the field $\exp({i\hbar_{\rm eff}P\xi/2})\Psi(Q+\xi/2,z)$. 
\par
  A polarizing beam-splitter divides the field {into two spatially identical components} and a Dove prism ($DP2$) placed inside 
  the interferometer realizes opposite 90$^\circ$ spatial rotations in the two counter-propagating {transverse spatial modes}, resulting in a total relative rotation by $180^\circ$. {The modes are recombined, and projected onto the diagonal polarization direction before detection by} an area-integrating ``bucket" detector. 
 The measured intensity is composed of three terms $I=I_1+I_2+I_{\rm int}$, where the sum $I_1+I_2$ is constant and equals {half} the total input intensity. The term $I_{\rm int}$ originates from the interference between the counter-propagating beams.  Due to the relative rotation implemented by $DP2$, $I_{\rm int}$ is proportional to: 
 \begin{equation}\label{eq:wigner-otica}
 \mathcal{W}(Q,P)=A\int_{-\infty}^\infty \Psi\left (Q+\frac{\xi}{2}\right )\Psi^*\left (Q-\frac{\xi}{2}\right )e^{i\hbar_{\rm eff}P\xi}d\xi,
\end{equation}
 The right-hand side of Eq.\eqref{eq:wigner-otica}
 is an integral over spatial variable $\xi$ of the overlap between the displaced field, $\exp({i\hbar_{\rm eff}P\xi/2})\Psi(Q+\xi/2,z)$, and its 
  complex conjugate with the transformation $\xi\rightarrow -\xi$.   {This transformation corresponds to the $180^\circ$ relative rotation  implemented by $DP2$.  It is important to note that slight polarization transformations introduced by the Dove prism~\cite{polarize-dove} inside the Sagnac interferometer does not alter significantly the measured Wigner function.}
  \par
In this way,  the {amplitude of the} optical Wigner function at point $(Q,P)$ can be obtained by measuring the intensity at the interferometer 
 output, {for different settings of the tilt angle and displacement of the steering mirror $M1$.}   In our experiment these parameters  were controlled {with high resolution} motorized stages.  At the exit of the interferometer we use a quarter wave plate (QWP) that is tilted to correct the polarization aberrations introduced by $DP2$ inside the interferometer \cite{polarize-dove}. 
 \par
\subsection*{Analysis of the experimental data}
The final state of the optical KHO is obtained at the output plane $z_0$, indicated in Fig. \ref{fig:setup}(b).  However, the optical wave-function propagates through free space and several linear optical elements before reaching the steering mirror at the entrance of the Sagnac interferometer, where the optical Wigner function is measured.  One must therefore take into account this evolution before comparing measurements with the theoretical Wigner functions, obtained from numerical calculation of the quantum KHO evolution. This is done by calculating the total linear transformation $\mathbb{M}$, resulting from {propagation through} all of the optical elements and free space between the output plane $z_0$ and the steering mirror shown in Fig. \ref{fig:setup} (b). 
\par
The expected Wigner function in the measurement plane can be written as 
\begin{equation}
W(x,p_x)=W_x^{\rm (KHO)}(x',p_x')\times W_y^{\rm(Gauss)}(y',p_y'), 
\end{equation}
where 
$\left(x'\;\;y'\;\;p_x'\;\;p_y'\right)^T=\mathbb{M}^{-1}\left(x\;\;0\;\;p_x\;\;0\right)^T$ {are the transformed coordinates}. The function 
$W_x^{\rm (KHO)}(x,p_x)$ represents the Wigner function due to KHO evolution of the initial Gaussian state {implemented in} the $x$ transverse spatial direction. The function  
$W_y^{\rm (Gauss)}(y,p_y)$ refers to the Gaussian state describing the $y$ transverse spatial direction at plane $z_0$. This spatial mode does not undergo KHO the evolution. 
\par
{Experimental errors associated to the propagation though the optical elements between the output plane $z_0$ and the steering
mirror results in} scaling, skewing and rotation of the final measured Wigner Function, 
$W^{\rm (exp)}(x,p_x)$, in comparison to {$W^{\rm (KHO)}(x,p_x)$}. {These uncertainties are due principally to errors in lens placement, misalignment of the optical elements, and diffraction.}  Nevertheless, a single linear transformation $\mathbb E$ corrects these errors, such that 
$W^{\rm (exp)}(x,p_x)=W{^{\rm (KHO)}}\left(\mathbb{E}(x,p_x)\right )$. Both $\mathbb M$ and $\mathbb E$ depend only on the experimental setup, and are the same for all measurements of Wigner 
functions, including the case in which the spatial light modulator is turned off. In this case, the implemented evolution is that of a simple harmonic oscillator (SHO).  {With the KHO turned off, we determined the value of $\mathbb E$, and used these values to correct all of the KHO Wigner functions.  This was repeated for each harmonic evolution used.}
\par
\subsection*{Implemention of a dephasing-type decoherence channel}
The spatial light modulator (SLM) imprints a phase on the horizontal polarization component of the light beam, but leaves  the phase of the vertical polarization component unchanged.  {Without the SLM, the optical system is designed to implement SHO evolution via the fractional Fourier transform systems composed of the cylindrical lens and free-space propagation.}   
Therefore, if diagonally polarized light is used in the KHO optical setup, one obtains the transformation: {\begin{eqnarray}|\Psi\rangle\otimes|+\rangle\;\rightarrow\; \frac{1}{\sqrt{2}}\left(\oper U_{\rm KHO} |\Psi\rangle\otimes|H\rangle+ \oper U_{\rm SHO} |\Psi\rangle\otimes|V\rangle \right) \label{transformation}\end{eqnarray}}
where $|+\rangle=(|H\rangle+|V\rangle)/\sqrt{2}$,
and $|\Psi\rangle$ designates the transverse spatial mode of the light beam. The evolution operators $\oper U_{\rm KHO}$ and $\oper U_{\rm SHO}$ act on the spatial degree of freedom, and correspond to the evolution of the KHO and the SHO, respectively. The total evolution given in Eq.(\ref{transformation}) could be interpreted as 
the quantum evolution of a qubit interacting (at the instant of the kick) with a quantum KHO via
a dephasing-type of coupling of the form $ -(s-s_H)K\cos Q/(s_V-s_H)  |H\rangle\langle H|$,
 where we denote with $s_H$ and $s_V$ the linear polarization eigenvalues of $|H\rangle$ and $|V\rangle$ respectively and $s=s_H$ or $s=s_V$. 
\par
Performing a polarization measurement of the output beam using a ``bucket" detector is equivalent to tracing over the spatial degree of freedom, and yields,
\begin{eqnarray}\oper\rho=\frac{1}{2}\left(|H\rangle\langle H| +|V\rangle\langle V| +f | H\rangle\langle V| +f^* | V\rangle\langle H| \right), 
\end{eqnarray}
where $f=\langle\Psi|\oper U_{\rm SHO}^\dagger\oper U_{\rm KHO}|\Psi\rangle$ is the overlap between the SHO and KHO quantum states. 
The purity of the polarization state is given by ${\rm Tr} \oper\rho^2 = (1+|f|^2)/2$, and decays with $f$ \cite{Lemos2011}.
\par 
We use wave plates and a polarizing beam splitter, to perform {quantum polarization state tomography} using the standard recipe \cite{NielsenChuang,james01} to obtain the density matrix $\rho$, from which we calculate the results shown in Fig. \ref{pure}.

\section*{Acknowledgements}
Financial support was provided by Brazilian agencies CNPq, CAPES, FAPERJ, and the Instituto Nacional de Ci\^encia e Tecnologia - Informa\c{c}\~ao Qu\^antica.  
\section*{Author Contributions}
G.L.,  S.W., P.R. and F.T. jointly conceived the experimental scheme and methodology.  G.L. and R.G. performed the experiments. All authors discussed the results and participated in the manuscript preparation.

\section*{Additional information}
{\bf Supplementary information} accompanies this paper at http://www.nature.com/ naturecommunications



\begin{thebibliography}{10}

\bibitem{stockmann}
St\"ockmann, H-J.
\newblock {\em Quantum Chaos, An introduction}.
\newblock (Cambridge University Press, Cambridge, 1999).

\bibitem{Izrailev}
Izrailev, F.~M.
\newblock Simple models of quantum chaos: spectrum and eigenfunctions.
\newblock {\em Phys. Rep.} {\bf196}, 299 -- 392, (1990).

\bibitem{steck2001}
Steck, D.A., Oskay, W.H.  \& Raizen, M.G.
\newblock {Observation of chaos-assisted tunneling between islands of
  stability}.
\newblock {\em Science} {\bf 293}, 274 (2001).


\bibitem{Hensinger2001}
Hensinger, W.K. \textit{et al.}
\newblock {Dynamical tunnelling of ultracold atoms.}
\newblock {\em Nature} {\bf 412}, 52 (2001).


\bibitem{raizen_pre1999}
 Bharucha, C.~F. \textit{et al.}
\newblock Dynamical localization of ultracold sodium atoms.
\newblock {\em Phys. Rev. E} {\bf 60}, 3881--3895 (1999).

\bibitem{kr-exp0}
Moore, F.~L. \textit{et al.}
\newblock Atom optics realization of the quantum $\delta{}$-kicked rotor.
\newblock {\em Phys. Rev. Lett.} {\bf 75}, 4598--4601 (1995).

\bibitem{nature-chaos}
Chaudhury, S. \textit{et al.}
\newblock Quantum signatures of chaos in a kicked top.
\newblock {\em Nature} {\bf 461}, 768--771 (2009).

\bibitem{Chabe2008}
Chab{\'e}, J. \textit{et al.}
\newblock {Experimental observation of the Anderson metal-insulator transition
  with atomic matter waves}.
\newblock {\em Phys. Rev. Lett.} {\bf 101}, 255702 (2008).

\bibitem{Sadgrove_review}
Sadgrove, M.  \& Wimberger, S.
\newblock A pseudoclassical method for the atom-optics kicked
  rotor: from theory to experiment and back.
\newblock In P.R.~Berman E.~Arimondo and C.C. Lin, editors, {\em Advances in
  Atomic, Molecular, and Optical Physics} {\bf 60},  315 -- 369 (Academic Press,
  2011).

\bibitem{Ryu2006}
Ryu, C. \textit{et al.}
\newblock High-order quantum resonances observed in a periodically kicked
  bose-einstein condensate.
\newblock {\em Phys. Rev. Lett.} {\bf 96}, 160403 (2006).

\bibitem{Fischer2000}
Fischer, B., Rosen, A., Bekker, A.  \&  Fishman, S.
\newblock Experimental observation of localization in the spatial frequency
  domain of a kicked optical system.
\newblock {\em Phys. Rev. E} {\bf 61}, 4694--4697 (2000).

\bibitem{Schwartz2007}
Schwartz, T.,  Bartal, G., Fishman, S.  \& Segev, M.
\newblock {Transport and Anderson localization in disordered two-dimensional
  photonic lattices}.
\newblock {\em Nature} {\bf 446}, 52--55 (2007).

\bibitem{zaslavsky}
Zaslavsky, G.~M.,  Sagdeev, R.~Z.,  Usikov, D.~A. \&  Chernikov, A.~A..
\newblock {\em Weak Chaos and Quasi-Regular Patterns}.
\newblock (Cambridge University Press, Cambridge, 1991).

\bibitem{fromhold}
 Fromhold, T.~M. \textit{et al.}
\newblock Effects of stochastic webs on chaotic electron transport in
  semiconductor superlattices.
\newblock {\em Phys. Rev. Lett.} {\bf 87}, 046803 (2001).

\bibitem{fromhold1}
 Fromhold, T.~M. \textit{et al.}
\newblock Chaotic electron diffusion through stochastic webs enhances current
  flow in superlattices.
\newblock {\em Nature} {\bf 428}, 726--730 (2004).

\bibitem{cirac-zoller}
 Gardiner, S.~A.,  Cirac, J.~I. \& Zoller, P.
\newblock Quantum chaos in an ion trap: The delta-kicked harmonic oscillator.
\newblock {\em Phys. Rev. Lett.} {\bf 79}, 4790--4793 (1997).

\bibitem{casati-chirikov}
Casati, G. \& Chirikov, B., editors.
\newblock {\em Quantum Chaos: Between Order and Disorder}.
\newblock (Cambridge University Press, Cambridge, 1995).

\bibitem{Schlunk2003}
Schlunk, S., d'Arcy, M., Gardiner, S. \& Summy, G.
\newblock Experimental observation of high-order quantum accelerator modes.
\newblock {\em Phys. Rev. Lett.} {\bf 90}, 124102 (2003).

\bibitem{Prange}
Prange, R.~W. \& Fishman, S.
\newblock {Experimental realizations of kicked quantum chaotic systems}.
\newblock {\em Phys. Rev. Lett.} {\bf 63}, 704--707 (1989).

\bibitem{berry}
Berry, M.~V., Balazs, N.~L., Tabor, M. \& Voros, A.
\newblock Quantum maps.
\newblock {\em Ann. Phys.} {\bf 122}, 26 -- 63 (1979).

\bibitem{baker-osorio}
 Hannay, J.~H.,  Keating, J.~P. \&  Ozorio~de Almeida, A.~M.
\newblock Optical realization of the baker's transformation.
\newblock {\em Nonlinearity} {\bf 7}, 1327--1342 (1994).

\bibitem{artuso_prl1992}
Artuso, R. \textit{et al.} \newblock {Phase diagram in the kicked Harper model}.
\newblock {\em Phys. Rev. Lett.} {\bf 69}, 3302--3305 (1992).

\bibitem{carvalho_buchleitner2004}
Carvalho, A. \& Buchleitner, A.
\newblock {Web-assisted tunneling in the kicked harmonic oscillator}.
\newblock {\em Phys. Rev. Lett.}  {\bf 93}, 04101 (2004).

\bibitem{Borgonovi1995}
Borgonovi, F. \& Rebuzzini, L.
\newblock Translational invariance in the kicked harmonic oscillator.
\newblock {\em Phys. Rev. E} {\bf 52}, 2302--2309 (1995).

\bibitem{stoler}
Stoler, D.
\newblock Operator methods in physical optics.
\newblock {\em J. Opt. Soc. Am.} {\bf 71}, 334--341
  (1981).

\bibitem{shamir82}
Nazarathy, M. \& Shamir, J.
\newblock First-order optics - a canonical operator representation: Lossless
  systems.
\newblock {\em J. Opt. Soc. Am. A} {\bf 72}, 356--364
  (1982).
\bibitem{nienhuis93}
Nienhuis, G. \& Allen, L.
\newblock Paraxial wave optics and harmonic oscillators.
\newblock {\em Phys. Rev. A} {\bf 48}, 656--665
  (1993).
  
\bibitem{ozaktas}
Ozaktas, H.~M., Zalevsky, Z. \& Apler Kutay, M.
\newblock {\em The Fractional Fourier Transform: with Applications in Optics
  and Signal Processing}.
\newblock (John Wiley and Sons Ltd., West Sussex, 2001).

\bibitem{bastiaans}
Bastiaans, M.~J.
\newblock The Wigner distribution function applied to optical signals and
  systems.
\newblock {\em Opt. Comm.} {\bf 25}, 26--30 (1978).

\bibitem{walmsley-wigner}
Mukamel, E.,  Benaszek, K.,  Walmsley, I.~A. \& Dorrer, C.
\newblock Direct measurement of the spatial Wigner function with
  area-integrated detection.
\newblock {\em Opt. Lett.} {\bf 28}, 1317--1319 (2003).

\bibitem{zurek-nature}
Zurek, W.~H.
\newblock Sub-Planck structure in phase space and its relevance to quantum
  decoherence.
\newblock {\em Nature} {\bf 412}, 712--717 (2001).

\bibitem{Maia2008}
Maia, R.~N.~P., Nicacio, F., Vallejos, R.~O. \& Toscano, F.
\newblock {Semiclassical propagation of gaussian wave packets}.
\newblock {\em Phys. Rev. Lett.} {\bf 100}, 184102 (2008).

\bibitem{Schubert2012}
Schubert, R., Vallejos, R.~O.\& Toscano, F.
\newblock {How do wave packets spread? Time evolution on Ehrenfest time
  scales}.
\newblock {\em J. Phys. A-Math. Theo.} {\bf 45}, 215307  (2012).

\bibitem{haug2005}
Haug, F. \textit{et al.}
\newblock Motional stability of the quantum kicked rotor: A fidelity approach.
\newblock {\em Phys. Rev. A} {\bf 71}, 043803 (2005).

\bibitem{gorin2006}
Gorin, T.,  Prosen, T.,  Seligman, T.~H. \& \v{Z}nidari\u{c}, M.
\newblock Dynamics of Loschmidt echoes and fidelity decay.
\newblock {\em Phys. Rep.} {435}, 33--156 (2006).

\bibitem{Lemos2011}
Lemos, G.~B. \& Toscano, F.
\newblock Decoherence, entanglement decay, and equilibration produced by
  chaotic environments.
\newblock {\em Phys. Rev. E} {\bf 84}, 016220 (2011).

\bibitem{NielsenChuang}
Nielsen, M.~A. \& Chaung, I.~L.
\newblock {\em Quantum Compuation and Quantum Information}.
\newblock (Cambridge University Press, Cambridge, 2000).

\bibitem{james01}
James, D.~F.~V., Kwiat, P.~G., Munro, W.~J \& White, A.~G.
\newblock Measurement of Qubits.
\newblock {\em Phys. Rev. A} {\bf 64}, 053212 (2001).

\bibitem{polarize-dove}
Moreno, I., Paez, G. \&  Strojnik, M.
\newblock Polarization transforming properties of dove prisms.
\newblock {\em Opt. Comm.}, {\bf 220}, 257--268 (2003).

\end{thebibliography}
\end{document}